\documentstyle[epsf,aps,pre]{revtex}
\input epsf
\begin{document}
\bibliographystyle{prsty}

\title{Rotating Convection in an Anisotropic System}
\author{Alex Roxin and Hermann Riecke}
\address{Engineering Science and Applied Mathematics,
Northwestern University, Evanston, Il 60208, USA}
\date{\today}
\maketitle
\begin{abstract}
We study the stability of patterns arising in rotating convection in weakly anisotropic systems using a modified 
Swift-Hohenberg equation.  The anisotropy, either an 
endogenous characteristic of the system or induced by external forcing, can stabilize periodic rolls in the K\"uppers-Lortz
chaotic regime.  For the particular case of rotating convection with time-modulated rotation where recently, in experiment,
 chiral patterns have been observed in otherwise K\"uppers-Lortz-unstable regimes, we show how the underlying base-flow breaks the 
isotropy, thereby affecting the linear growth-rate of convection rolls in such a way as to stabilize spirals and targets.  
Throughout we compare analytical results to numerical simulations of the Swift-Hohenberg equation.
\end{abstract}
\pacs{123}

\section{Introduction}

Pattern formation in thermal convection of a rotating fluid layer has been the subject of much experimental and theoretical work 
in recent years. The effect of the Coriolis force on the dynamics of thermal instabilities makes this system relevant for
both astrophysical and geophysical fluid dynamics, while the appearance of spatio-temporally chaotic dynamics near onset
make it an attractive candidate for detailed analytical and numerical investigations of the origin and behavior of chaotic complex
patterns.  

K\"uppers and Lortz \cite{KuLo69} determined that  for rotation rates greater than a critical value $\Omega >\Omega_{cr}$ steady 
convective roll patterns are unstable to another set of rolls oriented at an angle $\beta$ relative to the first.  These results were
confirmed and extended by Clever and Busse \cite{ClBu79} who also determined the dependence of $\Omega_{cr}$ and $\beta$
on the Prandtl number of the fluid.  In an infinite system, these dynamics are persistent due to isotropy.  Busse and Heikes
\cite{BuHe80} used this fact, and the closeness of $\beta$ to $\frac{\pi}{3}$ to derive three coupled amplitude equations, in
 which rolls switch cyclicly as they approach a heteroclinic orbit.  In real systems, small amplitude noise
perturbs this orbit, leading to nearly periodic switching of rolls.  In sufficiently large systems the switching becomes
incoherent in space and causes the development
of patches of rolls with different orientations.  The ensuing dynamics are chaotic, \cite{HuEc97}, \cite{HuPe98}.

In recent experiments on rotating convection  \cite{ThBa02}, Thompson, Bajaj and Ahlers investigated the effect of a temporal
modulation of the rotation rate on the K\"uppers-Lortz (KL) state.  They find that for sufficiently large modulation concentric roll
patterns (targets) as well as multi-armed spirals can be stabilized and replace the chaotic KL state.  Focusing
on the target pattern, they find that the rolls in these patterns drift radially inward and they measure the dependence of the 
drift velocity on modulation amplitude and frequency, mean rotation rate, and heating.  They point out, the modulation sets up an
oscillatory azimuthal mean flow, which tends to align rolls along that direction.  Since the alignment signles out a 
 specific orientation, it breaks
the isotropy of the system.  Motivated by these findings we therefore investigate here the effect of anisotropy on roll patterns in systems
 exhibiting KL chaos.

Within the framework of a suitably extended SH-model we first study the stability of straight rolls in systems with broken chiral 
symmetry (modeling the Coriolis force due to rotation) and with weak anisotropy.  We then use these analytical results to interpret simulation 
of this SH-model in a cylindrical geometry in which we obtain target and spiral patterns as seen in experiment.   

\section{The stability of rolls with anisotropy}

We study the effect of weak anisotropy on the K\"uppers-Lortz state in the following modified Swift-Hohenberg model,
\begin{equation}
\partial_{t}\psi = \mu\psi+\alpha^{2}(\hat{n}\cdot\nabla)^{2}\psi -(\nabla^{2}+1)^{2}\psi-\psi^{3}+\gamma\hat{k}\cdot\lbrack
\nabla\times\lbrack (\nabla\psi)^{2}\nabla\psi\rbrack\rbrack , \label{SH}
\end{equation}
where $\hat{n}$ is a director indicating the preferred orientation, and $\alpha$ gives the strength of this anisotropy.  We
retain the up-down (Boussinesq) symmetry ($\psi\to -\psi$ ) by including only odd terms in $\psi$, and include a  nonlinear gradient 
term that breaks the chiral symmetry.  The rotation rate is therefore measured by $\gamma$. 
  Similar
models have been systematically derived from the fluid equations, with \cite{PoPa97} and without \cite{NeFr93,MaRi02} mean flow 
effects, and have enjoyed widespread use, e.g. \cite{MiPe92,FaFr92,SaRi00,CrMe94}.  
We mean (\ref{SH}) to be a model
equation and are concerned with the {\it qualitative} effect of anisotropy on the K\"uppers-Lortz instability.

Focusing on the weakly nonlinear regime and assumimg the anisotropy to be weak, we take $\mu = \epsilon^{2}
\mu_{2}$ and $\alpha = \epsilon\alpha_{2}$ with $\epsilon\ll 1$.  To leading order in $\epsilon$ the system is therefore
isotropic.
To study the effect of the anisotropy on the KL-instability 
we consider the weakly nonlinear competition of two sets of rolls with relative angle $\beta$ with the ansatz,
\begin{equation}
\psi = \epsilon (A(\tau)e^{i(\cos(\theta)x+\sin(\theta)y)}+B(\tau)e^{i(\cos(\theta +\beta )x+\sin(\theta +\beta)y)}
+c.c.)+ h.o.t.
\end{equation}
Thus here we do not analyze all side-band instabilities.
To leading order the system is isotropic and $\theta$ is a free parameter.  The complex amplitudes $A$ and $B$
 evolve on the slow timescale $\tau = 
\epsilon t$.  For concreteness we take $\hat{n} = \hat{e}_{y}$.  At order $\epsilon^{3}$, 
a solvability condition yields
\begin{eqnarray}
\partial_{\tau}A &=& \mu_{2}A-\alpha^{2}\sin^{2}(\theta)A-3|A|^{2}A-(6+4\gamma\sin\beta\cos\beta )|B|^{2}A, \label{Amp:A}\\
\partial_{\tau}B &=& \mu_{2}B-\alpha^{2}\sin^{2}(\theta + \beta )B-3|B|^{2}B-(6-4\gamma\sin\beta\cos\beta )|A|^{2}B. \label{Amp:B}
\end{eqnarray}

We examine the stability of rolls of orientation $\theta$ with respect to a set of rolls oriented $\beta$ to the first set of rolls.
With $\alpha = 0$ (isotropic case), the absolute orientation of the rolls $\theta$ is irrelevant, and we find they become first
 unstable
to rolls with orientation
\begin{equation}
\beta_{KL} = 45^{o}
\end{equation}
for
\begin{equation}
\gamma\ge\gamma_{KL} = \frac{3}{2}.
\end{equation}

Introducing $\alpha\ne 0$ leads to a dependence of both $\beta_{KL}$ and $\gamma_{KL}$ on the obsolute orientation of the
rolls $\theta$.  The growth rates of the perturbations are given by
\begin{eqnarray}
\sigma_{A} &=& -2(\mu_{2}-\alpha^{2}\sin^{2}{\theta}), \label{eig:A}\\
\sigma_{B} &=& \mu_{2}(-1+\frac{4}{3}\gamma\sin{\beta}\cos{\beta})-\alpha (\sin^{2}{\theta +\beta}+\lbrack\frac{4}{3}
\gamma\sin{\beta}\cos{\beta}-2\rbrack\sin^{2}{\theta}).
\end{eqnarray}

\begin{figure}
\centerline{
\epsfxsize=3.2in 
\epsfbox{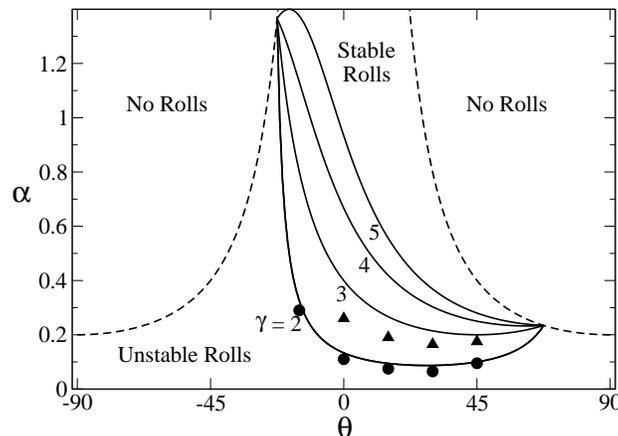}
}
\caption{Linear stability diagram of rolls with orientation $\theta$ in (\ref{Amp:A}, \ref{Amp:B}) with respect to rolls at a 
relative orientation of $\beta_{KL}$.  Here $\mu = 0.2$.  Numerical results are given by the solid symbols: triangles for
$\gamma = 3$ and circles for $\gamma = 2$.} \label{Fig:deltavstheta}
\end{figure}

As can be seen from (\ref{eig:A}) the anisotropy has shifted the onset of rolls with orientation $\theta$ to
\begin{equation}
\mu_{2}(\hat{\theta})=\alpha_{2}^{2}\sin^{2}{\hat{\theta}}. \label{onset}
\end{equation}
Thus rolls with orientation $\theta$ exist for $\mu_{2}>\mu_{2cr}(\theta)$.
For fixed $\mu$ this implies a neutral curve $\alpha(\theta)$ as shown by the dashed line in Figure \ref{Fig:deltavstheta}.
Rolls of orientation $\theta$ first become unstable to rolls of different orientation at 
\begin{equation}
\sigma_{B} \hspace{0.1in} = \hspace{0.1in}
\frac{\partial\sigma_{B}}{\partial\beta}  = 0.  \label{min}
\end{equation} 

Equation (\ref{min}) must be solved numerically for the linear stability limits.  
  Results are given in Figure \ref{Fig:deltavstheta} for various values of $\gamma$ (solid lines).
To test these stability results we perform numerical simulations in which we perturb straight rolls of orientation
$\theta$ by small-amplitude rolls of orientation $\theta +\beta$, where $\beta$ is chosen as the angle corresponding to
the maximal growth rate according to (\ref{min}).  To verify that no additional instabilities are present, we also perturb the 
rolls with small-amplitude noise.
  As can be seen from the solid symbols in Figure \ref{Fig:deltavstheta}, numerical simulations agree
well with the weakly nonlinear analysis for rotation rates $\gamma$ that are not too far above $\gamma_{KL}(\alpha)\sim 1.5$, for
which only weak anisotropy is needed for stability.
  For larger rotations rates $\gamma$,
the weakly nonlinear theory overestimates the amount of anisotropy $\alpha$ needed to stabilize rolls. For $\alpha = O(1)$ the anisotropy
affects the linear growth rate of rolls already in (\ref{SH}) and will introduce a significant dependence of the 
critical wavenumber 
on the orientation $\theta$.  Numerical results for 
larger $\alpha$ reveal that large amplitude rolls (with $\theta =0$) tend to grow and invade regions of rolls of
 other orientations front-wise.  In fact,
for $\alpha\to\infty$ only rolls with $\theta =0$ exist.

\begin{figure}[t!]
\centerline{
\epsfxsize=2.0in{\epsfbox{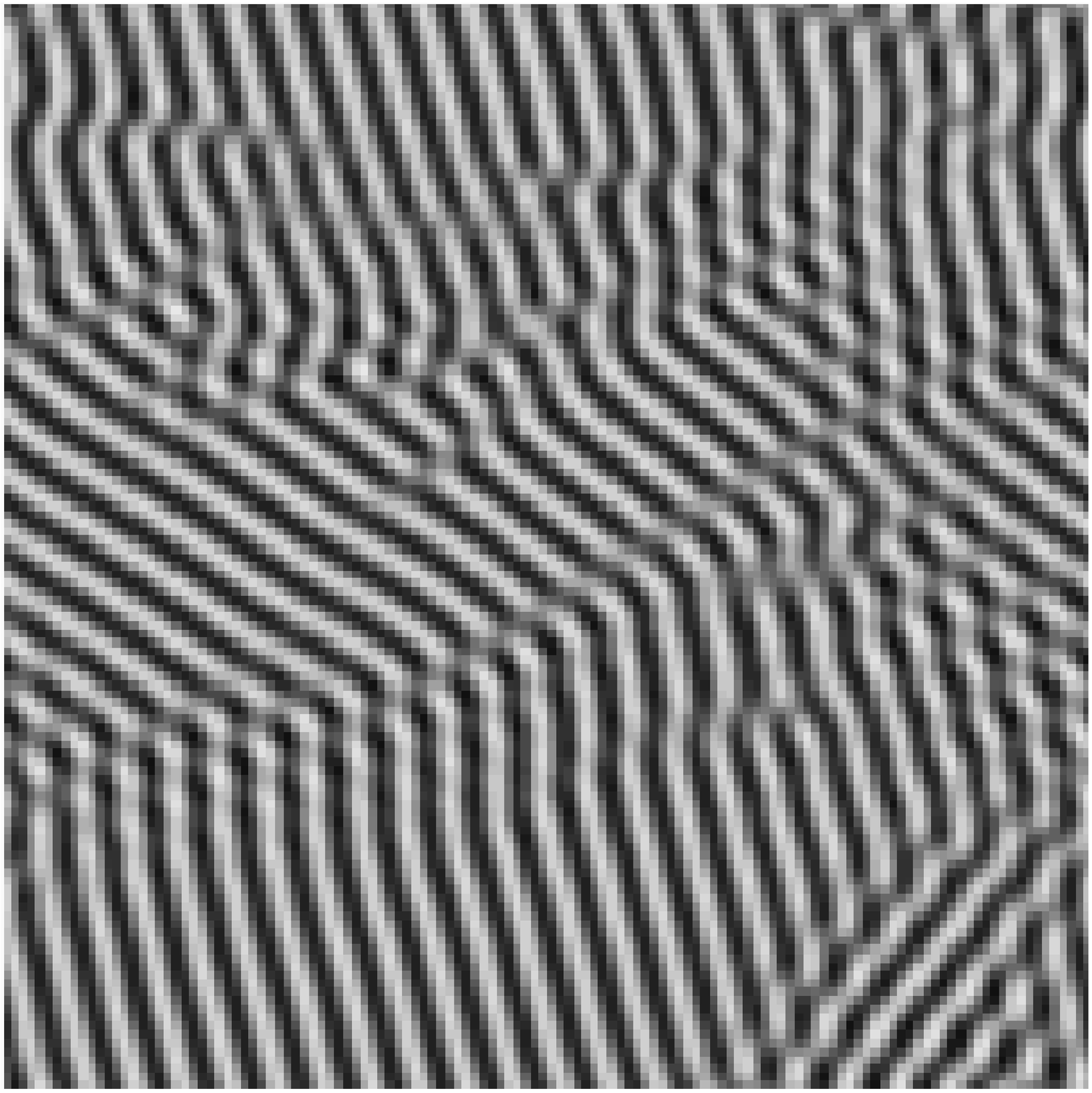}}
\hspace{0.1in}
\epsfxsize=2.0in{\epsfbox{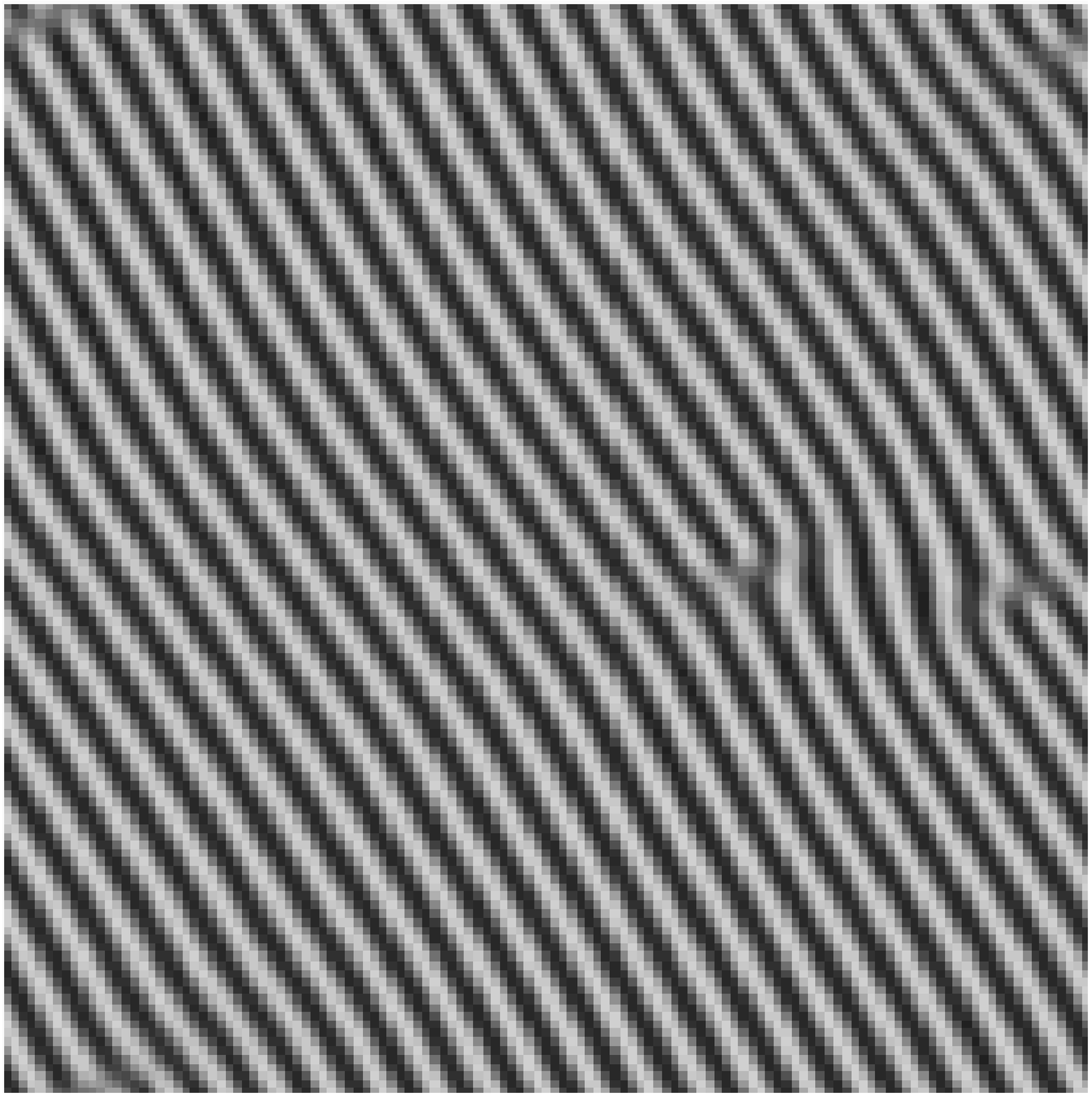}}}
\caption{Stabilization of rolls in the regime of domain chaos arising from the K\"uppers-Lortz instability.
a) A typical patch-work pattern of domain chaos, where the angle between patches
$\beta_{KL}=45$, and $\gamma = 2.0, \mu = 0.2, \alpha = 0.0$).  b) For the same values of the parameters with
$\alpha^{2} = 0.15$, rolls are stabilized.}
\end{figure}

Thus, weak anisotropy can stabilize periodic 
rolls arising in rotating convection in the K\"uppers-Lortz unstable regime.  Specifically, there is a finite band of angles 
$\theta$ with respect to the anisotropic director $\hat{n}$ such that rolls with this angle are stable to homogenous perturbations
of all possible orientations.  

\section{Modulated rotating convection: spirals and targets}

We now turn to the specific problem of rotating convection with periodically modulated rotation.  A thin layer of fluid of
height $d$ is 
heated from below and
bounded above and below by a rigid plate, which is rotated with an angular velocity $\Omega = \Omega_{o}(1+\delta\cos{\omega
 t})$.  For $\delta =0$ we recover the well-studied case of rotating convection \cite{KuLo69}-\cite{PoPa97}, 
\cite{MiPe92,FaFr92,Wo98}.  For $\delta\ll 1$ but nonzero, a 
nontrivial base flow is induced by the periodic motion of the rigid plates.  This flow advects
perturbations leading to thermal instabilities in such a way as to affect their growth rate.  Indeed, far from the axis of 
rotation, the onset of the
thermal instability is dependent on the orientation of periodic-roll perturbations with respect to the base flow in a manner
analogous to the linear operator in (\ref{SH}), \cite{Ke87}.  
Closer to the axis of rotation, the curvature of the base flow becomes 
significant and a straight-roll approximation is not a good one.

The dynamics are given by the Boussinesq fluid equations in a frame rotating at the mean angular velocity $\Omega_{0}$
\cite{KuLo69}-\cite{HuPe98} with rigid boundary conditions, which require the azimuthal velocity component at the 
top and bottom plates to obey,
\begin{equation}
u_{\theta} = \Re (\delta\Omega_{0}re^{i\omega t}).
\end{equation}
This condition induces an azimuthal shear flow, the strength of which grows with distance from the axis of rotation.
If we assume our flow is restricted to a finite geometry, Coriolis forces acting on this flow can be balanced by the 
radial pressure gradient as with the centrifugal force.

Far from the axis of rotation and, w.l.o.g., along the $x$-axis, the non-dimensionalized flow takes the form,
\begin{equation}
u_{\theta} = \delta\Re\Bigg(Pr\,\tau x\frac{\sinh{kz}-\sinh{k(z-1)}}{2\sinh{k}}e^{iPr\,\omega t}\Bigg), \label{base}
\end{equation}
where the plates are located at $z=0, 1$, $k = \sqrt{\frac{\omega}{2}}(1+i)$ and
\begin{equation}
Pr = \frac{\nu}{\kappa},\hspace{0.2in} \tau = \frac{2\Omega_{0}d^{2}}{\nu},
\end{equation}  
where the modulation frequency $\omega$ has been nondimensionalized with respect to the viscous diffusion time.
Note that (\ref{base}) satifies the equation of continuity, $\nabla\cdot\mathbf{u}=0$.

\begin{figure}[t!]
\centerline{
\epsfxsize=1.5in{\epsfbox{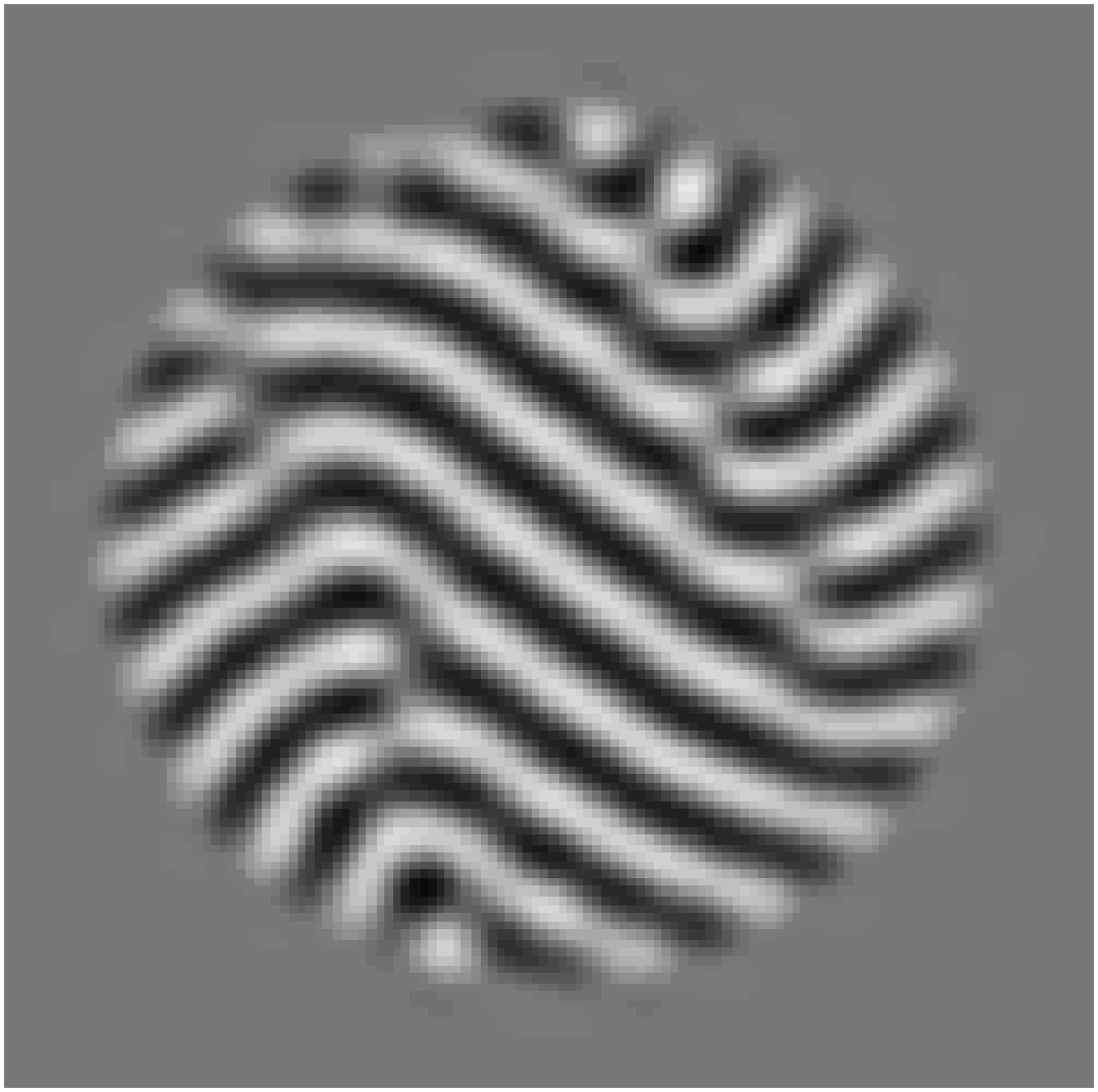}}
\hspace{0.1in}
\epsfxsize=1.5in{\epsfbox{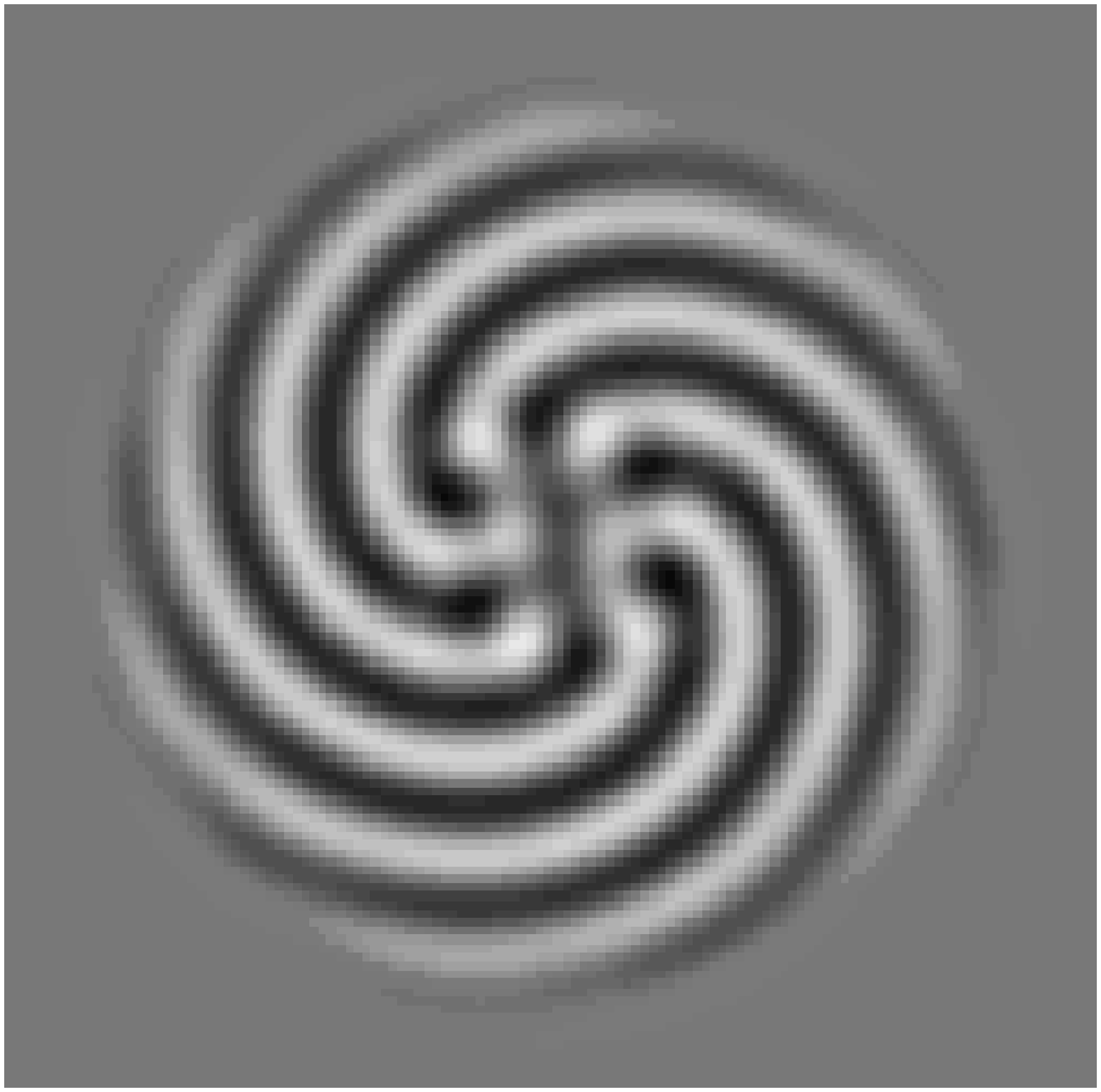}}
\hspace{0.1in}
\epsfxsize=1.5in{\epsfbox{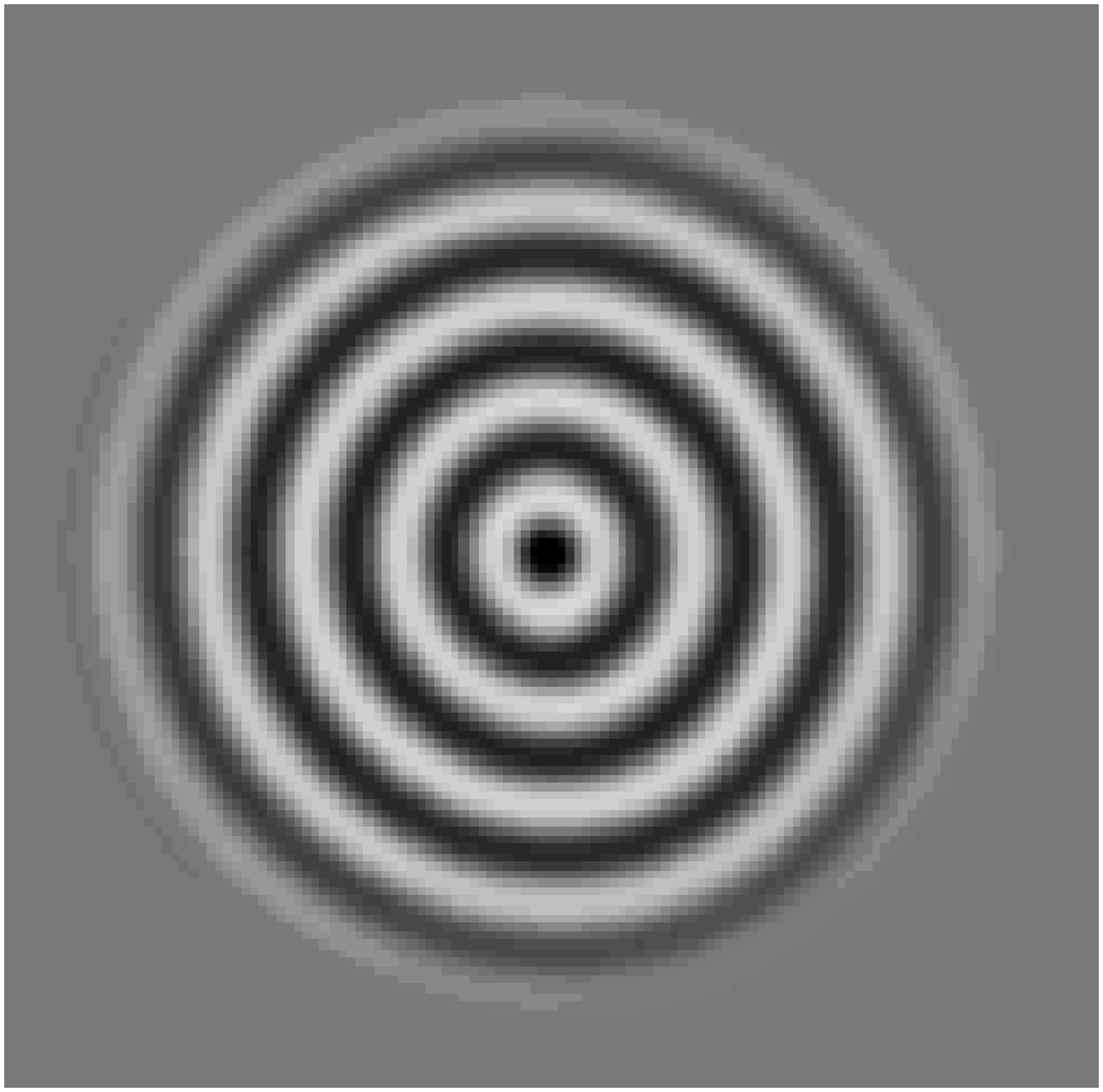}}}
\caption{Representative patterns: a) KL-state with $\delta = 0.0$ b) 6-armed spiral for $\delta = 0.0005$  
Equation (\ref{arms}) predicts six arms in this case. c) Target pattern for $\delta = 0.001$.  For all three
$\gamma = 2.0$, $\mu = 0.2$ and the system size $L=72$.}
\end{figure}
 
Thermal instabilities arising from the imposed temperature gradient are advected by (\ref{base}) and hence parametrically
forced with frequency $Pr\omega$.  Rayleigh-B\'enard convection has been studied
with time-modulated heating \cite{RoDa84,HoSw87,NiDo87,MeCa88}, and time-modulated gravity
 (parametrically accelerated) \cite{GrSa70}.  Here we note the following two facts which will allow us to deduce 
the effect of the base-flow (\ref{base}) on the instabilities to leading order.  Firstly, near onset,
the dynamics of the instability are slow compared with the period  of the oscillatory shear-flow for any finite 
rotation rate.  Secondly, the
forcing is invariant under the transformation $\delta\to -\delta, t\to t+\frac{\pi}{Pr\omega}$.  Averaging over the
fast oscillations with respect to the slow dynamics near onset, the base-flow affects the growth-rate of thermal
instabilities through mean-squared contributions (proportional to $\delta^{2}$).  
For the case (\ref{base}), the correction to the growth rate is of the 
form $u_{\theta}^{2}\partial^{2}_{y}$, whereas consideration of the full, cylindrical geometry yields $\frac{1}{r^{2}}u_{\theta}(r)^{2}
\partial^{2}_{\theta}$.  

With the abovementioned assumptions, we study the dynamics of patterns in (\ref{SH}) with,
\begin{equation}
\alpha^{2} = \delta^{2} r^{2}, \hspace{0.3in} \hat{n} = \hat{e}_{\theta}. \label{anis}
\end{equation}
We note that the scaling of the anisotropy (\ref{anis}) as linear in the distance from the axis of rotation, is only
correct far from the axis itself.  However, we retain this simplified form and hope to extract qualitatively correct
results.  In fact, simulations with other polynomial dependencies have revealed that only the monotonicity of the function
is important in determining qualitative features of the patterns.

Simulations reveal a wide variety of spiral patterns as well as targets.  For small $\delta$, where we expect the weakly-nonlinear
theory for periodic rolls to be valid sufficiently far from the core of the spiral, we are able to predict the number of spiral-arms
with reasonable accuracy.  Such an analysis can be understood from Figure \ref{Fig:deltavstheta}.  For a fixed 'rotation-rate'
$\gamma$, the strength of anisotropy $\alpha$ increases with distance from the core of the spiral.  There is thus a region in the
vicinity of the core where no rolls are stable, and rolls of a given orientation $\theta^{*}$ are selected at a distance $r^{*}$ as 
determined by condition (\ref{min}).  The number of resulting spiral-arms is then given by geometry as,
\begin{equation}
N = qr^{*}\sin{\theta^{*}}, \label{arms}
\end{equation}
where $q$ is the wavenumber of the rolls (arms).

Spirals or targets can be generated for the same parameter values given different initial conditions.  In general
an initial straight roll pattern will result in a target for sufficiently large $\delta$ whereas disordered initial
conditions generically yield spirals, even for strong anisotropy.

\begin{figure}[t!]
\centerline{
\epsfxsize=1.5in{\epsfbox{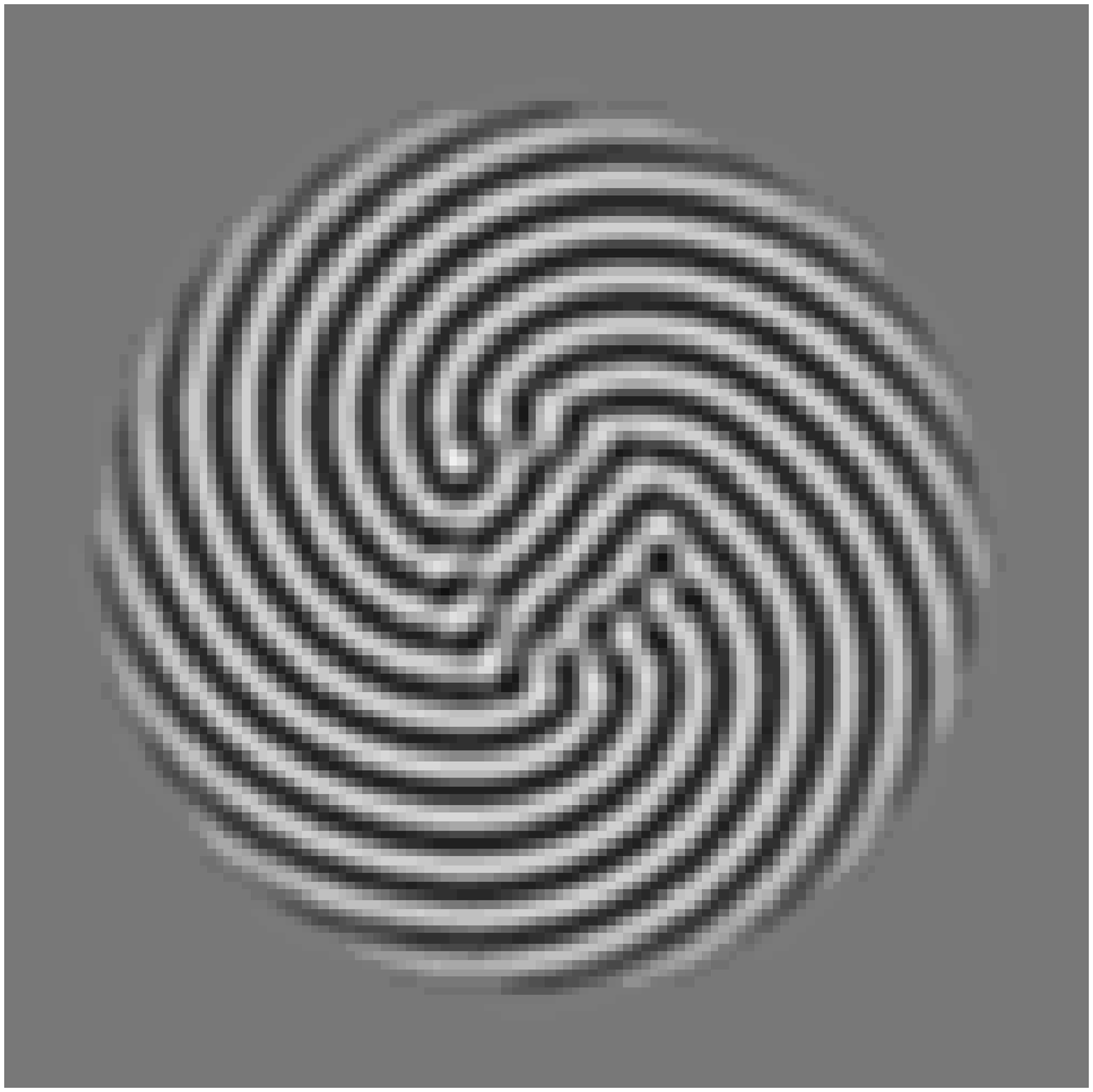}}
\hspace{0.1in}
\epsfxsize=1.5in{\epsfbox{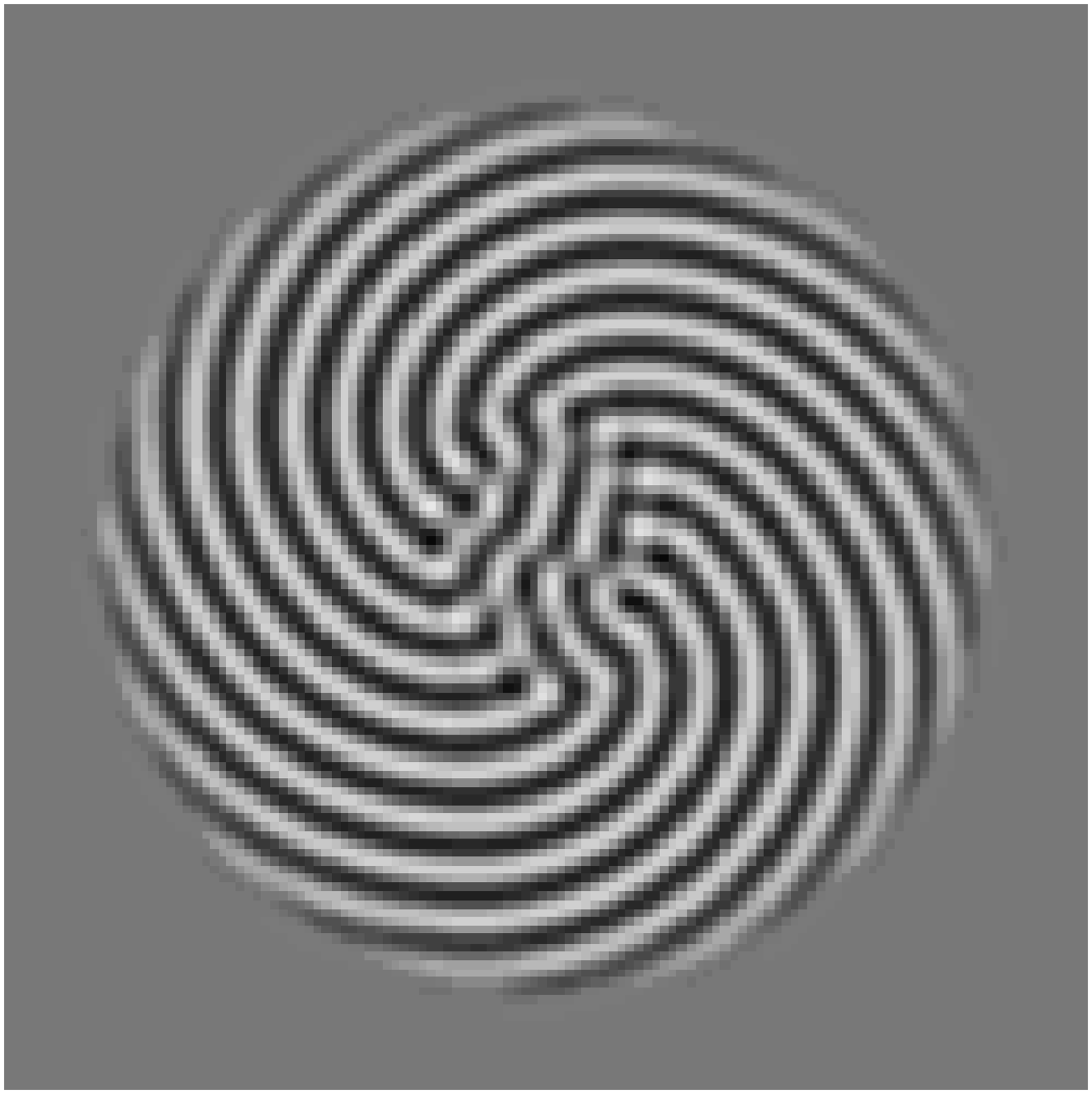}}
\hspace{0.1in}
\epsfxsize=1.5in{\epsfbox{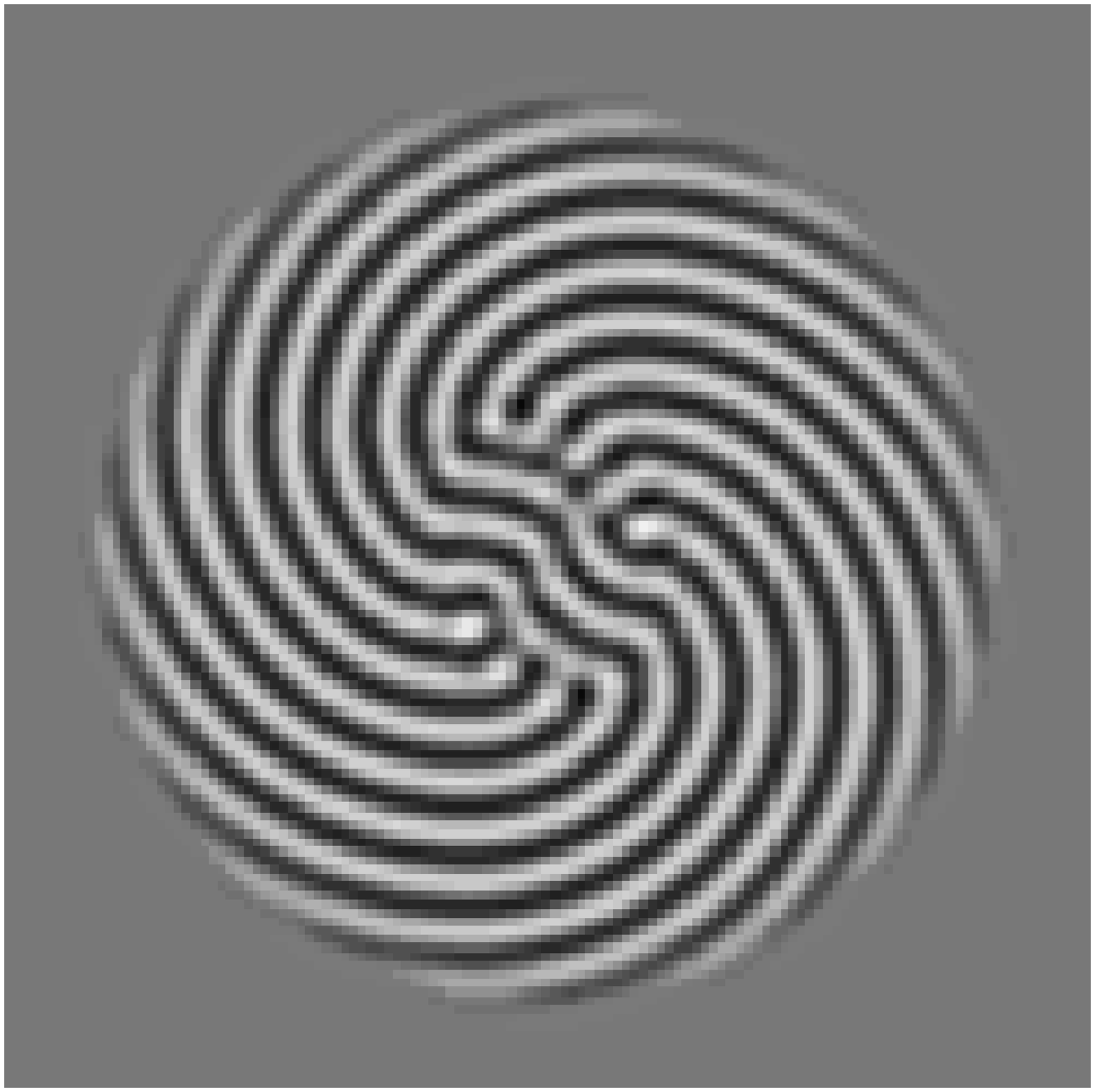}}
}
\caption{A stable spiral with a chaotic core.  The system size is $L=144$ with $\gamma = 2.0$, 
$\mu = 0.2$ and $\delta^{2} = 0.0001$.  Note that equation (\ref{arms}) predicts 14 arms.
  The core dynamics occur on a fast timescale w.r.t. the slow, solid-body rotation of the outer spiral.} \label{core}
\end{figure}

Interestingly, the orientation-selection mechanism given by (\ref{arms}) predicts the possibility of a large
region surrounding the core, within which no rolls are stable in the context of the weakly nonlinear theory.
If the anisotropy is sufficiently weak, one should see a disordered region of domain chaos, bounded by a 
stable spiral, given a large enough system.  Such a pattern is shown in Figure \ref{core}.

\section{Conclusion}
Spirals and targets arising in Rayleigh-B\'enard convection have been the subject of much theory and
experimental work,  \cite{PoPa97,HuEc97,FaFr92,HuEc94,EcHu95,HuEc93}.  Target patterns in low-Prandtl 
number convection are a consequence of horizontal, thermal gradients at the sidewalls of a 
cylindrical container, which tend to align rolls parallel to the walls \cite{HuEc93}.  Even with
sidewall forcing, the targets become unstable to straight rolls relatively close to threshold.  In 
rotating convection, the target patterns arising from such sidewall forcing undergo a mean drift 
\cite{FaFr92} due to the breaking of reflection symmetry by the applied rotation.  However, in the 
case of rotating convection with a modulated rotation rate, the chiral patterns are not a consequence
of the system geometry, but rather are induced by an isotropy-breaking shear flow, which acts 
azimuthally.  We have shown that these patterns are stable in regimes where one would see spatio-temporal 
chaos in the absence of modulation.  Our analysis indicates that the shear flow acts to 
stabilize rolls within a band of stable orientations w.r.t. the azimuthal flow itself.  This 
leads naturally to a chiral pattern.
 
The qualitative agreement between the types of patterns observed in experiment \cite{ThBa02} and those
studied here make the selection mechanism described in section III plausible.  Spirals and targets arise
through the interaction of the destabilizing process responsible for the KL instability and the 
stabilizing effect of the azimuthal mean flow (MF).  Quantitative comparison of the dependence of the 
pattern behavior on the reduced Rayleigh number, rotation rate, and amplitude and frequency of the modulation
with experiment is, however, not possible within the framework of (\ref{SH}).  However, qualitatively the genericity of the 
appearance of spirals patterns under modulated rotation with disordered intial conditions (KL state) seems
to hold both under experimental conditions and here.  Targets, on the other hand, 
 must be generated with care but then persist
as stable patterns over a wide range of parameter values.

The target patterns observed in \cite{ThBa02} travel inwards radially,
collapsing periodically at the center.  The origin of this drift has not been
identified yet. One possibility is that the time-periodic component of the
Coriolis force acting on the time-periodic azimuthal flow generates a radial
flow with a steady component, which would advect the axisymmetric roll
pattern \cite{Pepriv,Mopriv}. Another possibility is that the drift is due to a
mismatch between the wavenumber selected by the umbilicus \cite{PoMa81}
and that selected by the container side-wall \cite{CrDa83,HoKr85}. The
competing selected wavenumbers set up a wavenumber gradient that
induces a drift of the pattern \cite{KrBe82,PaRi91}. It should be possible to
distinguish between these two mechanisms by comparing the dynamics in 
systems of different aspect ratio.  For larger systems, the wavenumber
gradient induced by the incompatibility of the selected wavenumbers would be
weakened, while the effective strength of the radial flow would naturally be
stronger due to the greater Coriolis force at larger radii. Experiments in such 
larger systems would also be of interest in view of the prediction that in such 
systems the core of the spirals would exhibit chaotic dynamics of the K\"uppers-Lortz
type.

In our simulations of (\ref{SH}) no radial drift of the concentric rolls was found.
This is not unexpected since in the absence of the chiral-symmetry breaking
term proportional to $\gamma$ eq.(\ref{SH}) is variational and persistent
dynamics are ruled out. To obtain drift the variational character of the system
has to be  broken to a sufficient degree. This may require much larger
rotation rates $\gamma$ or the introduction of additional non-variational
terms. We have not pursued this further since simulations of extensions of
(\ref{SH}) with possibly also modified boundary conditions would not allow
any quantitative comparison with experiments and would therefore be of
limited use in identifying the dominant mechanism responsible for the drift. 
  
We wish to acknowledge helpful discussions with Guenter Ahlers, Vadim
Moroz, and Werner Pesch. This work was supported by the Engineering
Research Program of the Office of Basic Energy Sciences at the Department
of Energy (DE-FG02-92ER14303), by a grant from NSF (DMS-9804673),
and by a NSF-IGERT fellowship through grant DGE-9987577.


\begin{references}

\bibitem{KuLo69}
G. K\"uppers and D. Lortz, J. Fluid Mech. {\bf 35},  609  (1969).

\bibitem{ClBu79}
R. Clever and F. Busse, J. Fluid Mech. {\bf 94},  609  (1979).

\bibitem{BuHe80}
F. Busse and K. Heikes, Science {\bf 208},  173  (1980).

\bibitem{HuEc97}
Y. Hu, R. Ecke, and G. Ahlers, Phys. Rev. E {\bf 55},  6928  (1997).

\bibitem{HuPe98}
Y. Hu, W. Pesch, G. Ahlers, and R.~E. Ecke, Phys. Rev. E {\bf 58},  5821
  (1998).

\bibitem{ThBa02}
K.~L. Thompson, K.~M. Bajaj, and G. Ahlers, Phys. Rev. E  (submitted).

\bibitem{PoPa97}
Y. Ponty, T. Passot, and P. Sulem, Phys. Fluids {\bf 9},  67  (1997).

\bibitem{NeFr93}
M. Neufeld, R. Friedrich, and H. Haken, Z. Phys. B {\bf 92},  243  (1993).

\bibitem{MaRi02}
A. Mancho, H. Riecke, and F. Sain, preprint  .

\bibitem{MiPe92}
J. Mill\'an-Rodr\'{\i}guez {\it et~al.}, Phys. Rev. A {\bf 46},  4729  (1992).

\bibitem{FaFr92}
M. Fantz, R. Friedrich, M. Bestehorn, and H. Haken, Physica D {\bf 61},  147
  (1992).

\bibitem{SaRi00}
F. Sain and H. Riecke, Physica D {\bf 144},  124  (2000).

\bibitem{CrMe94}
M. Cross, D. Meiron, and Y. Tu, Chaos {\bf 4},  607  (1994).

\bibitem{Wo98}
R. Worthing, Phys. Lett. {\bf 237},  381  (1998).

\bibitem{Ke87}
R.~E. Kelly,  in {\em Physiochemical Hydrodynamics}, edited by D.~B. Spalding
  (1987), p.\ 65.

\bibitem{RoDa84}
M.~N. Roppo, S.~H. Davis, and S. Rosenblat, Phys. Fluids {\bf 27},  796
  (1984).

\bibitem{HoSw87}
P.~C. Hohenberg and J.~B. Swift, Phys. Rev. A {\bf 35},  3855  (1987).

\bibitem{NiDo87}
J.~J. Niemela and R.~J. Donnelly, Phys. Rev. Lett. {\bf 59},
  2431  (1987).

\bibitem{MeCa88}
C. Meyer {\it et~al.}, Phys.~Rev.~Lett. {\bf 61},  947  (1988).

\bibitem{GrSa70}
P.~M. Gresho and R.~L. Sani, J. Fluid Mech. {\bf 40 part 4},  783  (1970).

\bibitem{HuEc94}
Y. Hu, R. Ecke, and G. Ahlers, Phys. Rev. Lett. {\bf 72},  2191  (1994).

\bibitem{EcHu95}
R. Ecke, Y. Hu, R. Mainieri, and G. Ahlers, Science {\bf 269},  1704  (1995).

\bibitem{HuEc93}
Y. Hu, R. Ecke, and G. Ahlers, Phys. Rev. E {\bf 48},  4399  (1993).

\bibitem{Pepriv}
W. Pesch, priv. comm.  .

\bibitem{Mopriv}
V. Moroz, priv. comm.  .

\bibitem{PoMa81}
Y. Pomeau and P. Manneville, J. Phys. (Paris) {\bf 42},  1067  (1981).

\bibitem{CrDa83}
M.~C. Cross, P.~G. Daniels, P.~C. Hohenberg, and E.~D. Siggia, J.Fluid Mech.
  {\bf 127},  155  (1983).

\bibitem{HoKr85}
P. Hohenberg, L. Kramer, and H. Riecke, Physica D {\bf 15},  402  (1985).

\bibitem{KrBe82}
L. Kramer, E. Ben-Jacob, H. Brand, and M. Cross, Phys. Rev. Lett. {\bf 49},
  1891  (1982).

\bibitem{PaRi91}
H.-G. Paap and H. Riecke, Phys. Fluids A {\bf 3},  1519  (1991).




\end{references}
\end{document}